# РОБАСТНОЕ УПРАВЛЕНИЕ НАДВОДНЫМ СУДНОМ С АДАПТИВНОЙ КОМПЕНСАЦИЕЙ СИНУСОИДАЛЬНЫХ ВОЗМУЩЕНИЙ С НЕИЗВЕСТНЫМИ ПАРАМЕТРАМИ

О.И. Борисов, Ф.Б. Громова, А.Ю. Живицкий, А.А. Пыркин

Университет ИТМО, 197101, Санкт-Петербург, Россия
E-mail: borisov@itmo.ru

В работе решена задача удержания заданного положения надводного судна в условиях действия синусоидальных возмущений с неизвестными параметрами. Предлагаемый алгоритм управления основан на геометрическом подходе с использованием адаптивной внутренней модели и расширенного наблюдателя. В работе аналитически доказана ограниченность траекторий системы и их полуглобальная сходимость к сколь угодно малому множеству. Работоспособность алгоритма проиллюстрирована компьютерным моделированием.

*Ключевые слова*: *компенсация возмущений, геометрический подход, адаптивная внутренняя модель, расширенный наблюдатель, надводное судно.*

**Введение.** Задача управления в условиях действия синусоидальных возмущений является одной из классических в теории систем управления. Ее решению посвящен ряд научных работ (см. [1-5]). С ней также тесно связана задача параметрической идентификации синусоидальных сигналов и, в частности, оценивание частот (см. [6, 7]).

В работе [1] предложен алгоритм управления по выходу одномерным параметрически неопределенным линейным объектом, подверженного влиянию синусоидального возмущения с неизвестными параметрами. Предложенный подход предполагает двухэтапный синтез закона управления: стабилизация невозмущенного объекта с помощью регулятора с высоким коэффициентом усиления и собственно компенсация возмущения. В развитие этого результата в работе [3] приведено решение более сложной задачи, в которой канал управления характеризуется временным запаздыванием.

В работе [2] решена задача компенсации мультисинусоидального возмущения, действующего на нелинейный одномерный объект в форме Лурье с запаздыванием в канале управления. Нелинейная функция и параметры модели объекта предполагаются известными, а характеристический полином передаточных функций объекта по управлению, возмущению и нелинейности является гурвицевым. Предложенное решение основано на идентификационном подходе и упреждающей оценке периодических функций.

В настоящей работе развиваются методы управления в условиях возмущений с неизвестными параметрами. Предлагаемый алгоритм основан на геометрическом подходе и представляет собой адаптивную внутреннюю модель, объединенную с расширенным наблюдателем. Настоящая работа развивает результат, ранее опубликованный в [4], где решена задача управления надводным судном с компенсацией возмущений с известными частотами. Методика синтеза адаптивной внутренней модели для рассматриваемой задачи основана на методе, представленном в [5] для многоканальных систем в общем виде.

**Постановка задачи**. Рассмотрим динамическую модель надводного судна (см. главу 7 в [8])

$$M\ddot{x} + D\dot{x} = \tau + \tau_{\mathrm{d}}, \tag{1}$$

где $x = (X \quad Y \quad \Phi) \in \mathbb{R}^3$ — вектор положения и ориентации относительно параллельной системы координат надводного судна, $\dot{x} = (\dot{X} \quad \dot{Y} \quad \dot{\Phi}) \in \mathbb{R}^3$ — вектор скоростей, $\ddot{x} = (\ddot{X} \quad \ddot{Y} \quad \ddot{\Phi}) \in \mathbb{R}^3$ — вектор ускорений, $\tau = (\tau_1 \quad \tau_2 \quad \tau_3) \in$ $\mathbb{R}^3$ — вектор сил и моментов сил, развиваемых исполнительными приводами, $\tau_d = (\tau_{d1} \quad \tau_{d2} \quad \tau_{d3}) \in \mathbb{R}^3$ — вектор внешних сил и моментов сил, обусловленных влиянием волн, ветра и течения, $M \in \mathbb{R}^{3 \times 3}$ и $D \in \mathbb{R}^{3 \times 3}$ — матрицы соответственно инерции и демпфирования, содержащие неопределенные параметры.

Динамическая модель внешних возмущений $\tau_d$ описывается генератором вида

$$\begin{aligned} \dot{w} &= S(\varrho)w, \\ \tau_d &= Hw, \end{aligned} \qquad (2)$$

где $S(\varrho) = \mathrm{diag}\big(S_1(\varrho_1), S_2(\varrho_2), S_3(\varrho_3)\big) \in \mathbb{R}^{n_d \times n_d}$ — матрица состояния, характеризующаяся неопределенностью $\varrho = \{\varrho_1, \varrho_2, \varrho_3\}$ и $H = \mathrm{diag}(H_1, H_2, H_3) \in \mathbb{R}^{3 \times n_d}$ — матрица выхода.

Целью является синтез регулятора, обеспечивающего ограниченность траекторий системы и для любого сколь угодно малого $\epsilon > 0$ выполнение целевого условия

$$\lim_{t \to \infty} \|x_e(t)\| \leq \epsilon, \qquad (3)$$

где $x_e(t) = x(t) - x_r$ — вектор рассогласования между выходом $x(t)$ и постоянным задающим воздействием $x_r$.

**Основной результат.** Введем вектор рассогласования

$$\xi = \begin{pmatrix} \xi_1 \\ \xi_2 \end{pmatrix} = \begin{pmatrix} x_e \\ \dot{x}_e \end{pmatrix}$$

и представим систему (1) в виде модели ошибки как

$$\begin{aligned} \dot{\xi} &= A\xi + B\tau + Pw, \\ x_e &= C_e \xi, \end{aligned} \qquad (4)$$

где

$$A = \begin{pmatrix} 0_{3\times3} & I_{3\times3} \\ 0_{3\times3} & -M^{-1}D \end{pmatrix}, B = \begin{pmatrix} 0_{3\times3} \\ M^{-1} \end{pmatrix}, P = \begin{pmatrix} 0_{3\times3} \\ M^{-1}H \end{pmatrix}, C_e = (I_{3\times3} \quad 0_{3\times3}).$$

Известно [9], что для решения задачи управления по выходу необходимо проанализировать ее поведение в установившемся режиме и найти пару $\Pi, \Psi$, являющуюся решением выражений

$$\Pi S(\varrho) = A\Pi + B\Psi + P,$$
$$0 = C_e \Pi,$$

где $\Pi$ — значение вектора состояния в установившемся режиме, $\Psi$ — значение вектора управления в установившемся режиме.

Введем новые переменные $z = (z_1^T \ z_2^T)^T$ как

$$\begin{pmatrix} z_1 \\ z_2 \end{pmatrix} = \begin{pmatrix} I & 0 \\ T_1 & T_2 \end{pmatrix} \begin{pmatrix} \xi_1 \\ \xi_2 \end{pmatrix},$$

где $T_1$ и $T_2$ — обратимые матрицы, которые будут определены позднее, и преобразуем систему (4) как

$$\begin{pmatrix} \dot{z}_1 \\ \dot{z}_2 \end{pmatrix} = \begin{pmatrix} A_{11} & A_{12} \\ A_{21} & A_{22} \end{pmatrix} \begin{pmatrix} z_1 \\ z_2 \end{pmatrix} + \begin{pmatrix} 0_{3 \times 3} \\ B_2 \end{pmatrix} \tau + \begin{pmatrix} 0_{3 \times n_d} \\ P_2 \end{pmatrix} w, \qquad (5)$$

где

$$A_{11} = -T_2^{-1} T_1, \qquad A_{12} = T_2^{-1}, \qquad P_2 = T_2 M^{-1} H,$$
$$A_{21} = -(T_1 - T_2 M^{-1} D) T_2^{-1} T_1, \quad A_{22} = (T_1 - T_2 M^{-1} D) T_2^{-1} \quad B_2 = T_2 M^{-1},$$

причем матрица $T_2$ такая, что $B_2 > 0$.

Для компенсации внешних возмущений дополним систему (5) внутренней моделью вида

$$\begin{aligned} \dot{\eta} &= F\eta + G[\Gamma(\varrho)\eta + z_2 + \tau_0], \\ \tau &= -\Gamma(\varrho)\eta - z_2 - \tau_0, \end{aligned} \qquad (6)$$

где $F = \mathrm{diag}(F_1, F_2, F_3) \in \mathbb{R}^{n_d \times n_d}$ и $F_i$ — гурвицева матрица во фробениусовой форме, $G = \mathrm{diag}(G_1, G_2, G_3) \in \mathbb{R}^{n_d \times 3}$ и $G_i = (0 \ 0 \ \ldots \ 0 \ 1)^T$, $\Gamma(\varrho) = \mathrm{diag}\big(\Gamma_1(\varrho_1), \Gamma_2(\varrho_2), \Gamma_3(\varrho_3)\big) \in \mathbb{R}^{3 \times n_d}$ и $\Gamma_i(\varrho_i)$ вектор такой, что собственные числа матриц $F_i + G_i \Gamma_i(\varrho_i)$ и $S_i$ совпадают, $i = \{1,2,3\}$, $\tau_0 \in \mathbb{R}^3$ — управляющее воздействие, которое будет определено позднее.

Объединяя модель объекта (5) и внутреннюю модель (6), получим агрегированную систему

$$\begin{pmatrix}\dot{\eta}\\ \dot{z}_1\\ \dot{z}_2\end{pmatrix}=\begin{pmatrix}F+G\varGamma(\varrho) & 0_{n_d\times 3} & G\\ 0_{3\times n_d} & A_{11} & A_{12}\\ -B_2\varGamma(\varrho) & A_{21} & A_{22}-B_2\end{pmatrix}\begin{pmatrix}\eta\\ z_1\\ z_2\end{pmatrix}+\begin{pmatrix}G\\ 0\\ -B_2\end{pmatrix}\tau_0+\begin{pmatrix}0_{n_d\times n_d}\\ 0_{3\times n_d}\\ P_2\end{pmatrix}w. \quad (7)$$

Известно [10, 11], что существует матрица $\varSigma=\mathrm{diag}(\varSigma_1,\varSigma_2,\varSigma_3)\in\mathbb{R}^{n_d\times n_d}$, удовлетворяющая условиям

$$\varSigma S=\bigl(F+G\varGamma(\varrho)\bigr)\varSigma, \quad H=\varGamma(\varrho)\varSigma.$$

Чтобы показать, что в установившемся режиме внутренняя модель обеспечивает компенсацию внешних возмущений, введем новую переменную

$$\tilde{\eta}=\eta-\varSigma w,$$

дифференцируя которую, получим

$$\dot{\tilde{\eta}}=F\tilde{\eta}+G(\varGamma(\varrho)\tilde{\eta}+z_2-\tau_0),$$

а также выберем

$$\tau_0=-\varGamma\tilde{\eta},$$

с учетом этого преобразуем систему (7) как

$$\begin{pmatrix}\dot{\tilde{\eta}}\\ \dot{z}_1\\ \dot{z}_2\end{pmatrix}=\begin{pmatrix}F & 0_{n_d\times 3} & G\\ 0_{3\times n_d} & A_{11} & A_{12}\\ 0_{3\times n_d} & A_{21} & A_{22}-B_2\end{pmatrix}\begin{pmatrix}\tilde{\eta}\\ z_1\\ z_2\end{pmatrix}, \quad (8)$$

где видно, что действие внешних возмущений скомпенсировано.

Можно показать, что система (8) является глобально асимптотически устойчивой при выборе

$$T_1=MK_1, \quad T_2=MK_2-D, \quad K_1,K_2>0,$$

что соответствует номинальной линеаризации по входу-выходу системы (4) с помощью номинального закона управления

$$\tau=-\varGamma(\varrho)\eta-M\bigl(K_1\xi_1+K_2\xi_2+q(\xi_2)\bigr)-\varGamma\tilde{\eta},$$

где $q(\xi_2)=-M^{-1}D\xi_2$.

Поскольку матрица состояния генератора (2) характеризуется неопределенностью, заменим во внутренней модели (6) вектор $\varGamma(\varrho)$, зависимый от неопределенного параметра $\varrho$, на его оценку $\hat{\varGamma}$ и получим адаптивную внутреннюю модель

$$\begin{aligned}\dot{\eta} &= F\eta + G[\hat{\Gamma}\eta + z_2], \\ \tau &= -\hat{\Gamma}\eta - z_2.\end{aligned} \qquad (9)$$

Введем ошибку оценивания $\tilde{\Gamma} = \hat{\Gamma} - \Gamma(\varrho)$ и вектор $x = (\dot{\tilde{\eta}}^T \ \dot{\xi}_1^T \ \dot{\xi}_2^T)^T$, с учетом чего замкнутая система (8) примет вид

$$\begin{aligned}\dot{x} &= Ax + B\tilde{\Gamma}\eta, \\ z_2 &= Cx,\end{aligned} \qquad (10)$$

где

$$A = \begin{pmatrix} F & 0_{n_d \times 3} & G \\ 0_{3 \times n_d} & A_{11} & A_{12} \\ 0_{3 \times n_d} & A_{21} & A_{22} - B_2 \end{pmatrix}, B = \begin{pmatrix} G \\ 0_{3 \times 3} \\ -B_2 \end{pmatrix}, C = \begin{pmatrix} 0_{3 \times n_d} & 0_{3 \times 3} & I_{3 \times 3} \end{pmatrix},$$

причем матрица $A$ является гурвицевой.

Рассмотрим следующее утверждение (см. [5]) применительно к системе (10).

**Утверждение 1**. *Существует положительно определенная симметричная матрица $P$ такая, что*

$$PA + A^T P < 0, \quad PB = -C^T.$$

**Доказательство.** Выберем обратимую матрицу

$$T = \begin{pmatrix} I_{n_d \times n_d} & 0_{n_d \times 3} & GB_2^{-1} \\ 0_{3 \times n_d} & I_{3 \times 3} & 0_{3 \times 3} \\ 0_{3 \times n_d} & 0_{3 \times 3} & I_{3 \times 3} \end{pmatrix},$$

для которой выполняется

$$TB = \begin{pmatrix} 0_{n_d \times 3} \\ 0_{3 \times 3} \\ -B_2 \end{pmatrix}. \qquad (11)$$

После несложных вычислений

$$\tilde{A} = TAT^{-1}$$

получим

$$\tilde{A} = \begin{pmatrix} F & GB_2^{-1}A_{21} & -FGB_2^{-1} + GB_2^{-1}A_{22} \\ 0_{3 \times n_d} & A_{11} & A_{12} \\ 0_{3 \times n_d} & A_{21} & A_{22} - B_2 \end{pmatrix}.$$

Поскольку матрица $F$, являющаяся блочным элементом матрицы $\tilde{A}$, гурвицева по построению, то существует положительно определенная симметричная матрица $P_1$ такая, что

$$P_1 F + F^T P_1 < 0.$$

Рассмотрим положительно определенную матрицу

$$\tilde{P} = \begin{pmatrix} P_1 & 0_{n_d \times 3} & 0_{n_d \times 3} \\ 0_{3 \times n_d} & I_{3 \times 3} & 0_{3 \times 3} \\ 0_{3 \times n_d} & 0_{3 \times 3} & B_2^{-1} \end{pmatrix},$$

удовлетворяющую условию

$$\tilde{P}\tilde{A} + \tilde{A}^T \tilde{P} < 0,$$

с учетом чего определим

$$P = T^T \tilde{P} T$$

и получим

$$PA + A^T P = T^T [\tilde{P}\tilde{A} + \tilde{A}^T \tilde{P}] T < 0.$$

С учетом (11) заметим, что

$$PB = T^T \tilde{P} T B = T^T \tilde{P} \begin{pmatrix} 0_{n_d \times 3} \\ 0_{3 \times 3} \\ -B_2 \end{pmatrix} = \begin{pmatrix} 0_{n_d \times 3} \\ 0_{3 \times 3} \\ -I_{3 \times 3} \end{pmatrix} = -C^T,$$

что и требовалось доказать.

Рассмотрим функцию Ляпунова вида

$$V(x, \tilde{\Gamma}) = x^T P x + \sum_{i=1}^{3} \tilde{\Gamma}_i Q_i^{-1} \tilde{\Gamma}_i^T, \quad i = \{1,2,3\}, \tag{12}$$

где $Q = Q^T > 0$, $\tilde{\Gamma}_i = \hat{\Gamma}_i - \Gamma_i(\varrho_i)$.

Заметим $\dot{\hat{\Gamma}}_i = \dot{\tilde{\Gamma}}_i$ и обозначим закон адаптации как

$$\dot{\hat{\Gamma}}_i^T = \varphi_i(\cdot). \tag{13}$$

Дифференцируя (12), получим

$$\begin{aligned}\dot{V}(x,\tilde{\Gamma}) &= x^T[PA + A^TP]x + 2x^TPB\tilde{\Gamma}\eta + 2\sum_{i=1}^{3}\tilde{\Gamma}_i Q_i^{-1}\varphi_i(\cdot) \\ &= x^T[PA + A^TP]x - 2x^TC^T\tilde{\Gamma}\eta + 2\sum_{i=1}^{3}\tilde{\Gamma}_i Q_i^{-1}\varphi_i(\cdot) \\ &= x^T[PA + A^TP]x - 2z_2^T\tilde{\Gamma}\eta + 2\sum_{i=1}^{3}\tilde{\Gamma}_i Q_i^{-1}\varphi_i(\cdot),\end{aligned}$$

откуда, обозначив $z_2 = (z_{21}\ z_{22}\ z_{23})^T$, выберем

$$\varphi_i(\cdot) = Qz_{2i}\eta_i, \quad i = \{1,2,3\}, \tag{14}$$

с учетом чего получим

$$\dot{V}(x,\tilde{\Gamma}) = x^T[PA + A^TP]x \leq 0,$$

откуда следует, что все траектории ограничены и ошибки регулирования глобально асимптотически сходятся к нулю, несмотря на неопределенность вектора $\varrho$.

Однако адаптивная внутренняя модель (9), (13), (14) нереализуема, поскольку сигнал $z_2$ неизмерим. В связи с этим добавим в систему наблюдатель сигнала $z_2$, на базе которого получим реализуемую адаптивную внутреннюю модель вида

$$\begin{aligned}\dot{\eta} &= F\eta + G[\hat{\Gamma}\eta + \hat{z}_2], \\ \tau &= -\hat{\Gamma}\eta - \hat{z}_2,\end{aligned} \tag{15}$$

где $\hat{z}_1$ — оценка сигнала $z_1$, полученная с помощью алгоритма

$$\hat{z}_2 = \text{sat}_L \bar{z}_2 = \text{sat}_L \bar{M}[\sigma + K_1\hat{\xi}_1 + K_2\hat{\xi}_2], \tag{16}$$

где $\text{sat}_L(\cdot)$ — гладкая функция насыщения с пределом $L > 0$, $\bar{M} \in \mathbb{R}^{3\times 3}$ — обратимая матрица, удовлетворяющая условию

$$\|\,[M^{-1} - \bar{M}^{-1}]\bar{M}\,\|_1 \leq \delta_0 < 1,$$

$\hat{\xi}$ и $\sigma$ — элементы состояния расширенного наблюдателя вида

$$\begin{aligned}\dot{\hat{\xi}}_1 &= \hat{\xi}_2 + \kappa C_2(x_e - \hat{\xi}_1), \\ \dot{\hat{\xi}}_2 &= \sigma - \bar{M}^{-1}\hat{z}_2 + \kappa^2 C_1(x_e - \hat{\xi}_1), \\ \dot{\sigma} &= \kappa^3 C_0(x_e - \hat{\xi}_1),\end{aligned} \tag{17}$$

где $\kappa$ — высокий коэффициент усиления, $C_0, C_1, C_2 \in \mathbb{R}^{3\times 3}$ — положительно определенные матрицы такие, что собственные числа составной матрицы

$$A_o = \begin{pmatrix} -C_2 & I_{3\times 3} & 0_{3\times 3} \\ -C_1 & 0_{3\times 3} & I_{3\times 3} \\ -C_0 & 0_{3\times 3} & 0_{3\times 3} \end{pmatrix}$$

являются вещественными и отрицательными.

Обозначив $\hat{z}_2 = (\hat{z}_{21} \; \hat{z}_{22} \; \hat{z}_{23})^T$, заменим (14) на реализуемый закон адаптации вида

$$\varphi_i(\cdot) = Q\hat{z}_{2i}\eta_i, \quad i = \{1,2,3\}. \tag{18}$$

Введем ошибки оценивания $e = (e_1^T \; e_2^T \; e_3^T)^T$ как

$$\begin{aligned} e_1 &= \kappa^2(\xi_1 - \hat{\xi}_1), \\ e_2 &= \kappa(\xi_2 - \hat{\xi}_2), \\ e_3 &= q(\xi_2) + [\bar{M}^{-1} - M^{-1}]\hat{z}_2 - \sigma, \end{aligned}$$

дифференцируя которые получим модель ошибки оценивания

$$\dot{e} = \kappa[A_o - B_{o2}\Delta_0(\xi, e)C_o]e + B_{o1}\Delta_1(\xi, e) + B_{o2}\Delta_2(\xi, e),$$

где

$$A_o = \begin{pmatrix} -C_2 & I_{3\times 3} & 0_{3\times 3} \\ -C_1 & 0_{3\times 3} & I_{3\times 3} \\ -C_0 & 0_{3\times 3} & 0_{3\times 3} \end{pmatrix}, \quad B_{o1} = \begin{pmatrix} 0 \\ I \\ 0 \end{pmatrix}, \quad B_{o2} = \begin{pmatrix} 0 \\ 0 \\ I \end{pmatrix}, \quad C_o = (C_0 \; 0 \; 0),$$

а также известно, что существуют $\delta_0$ и $\delta_1$ такие, что

$$\|\Delta_0(\xi, e)\| \leq \delta_0 < 1, \quad \|\Delta_1(\xi, e)\| \leq \delta_1 \|e\|$$

для всех $(\xi, e)$ и $\kappa$.

Кроме того, для каждого $R > 0$ существует число $N$ такое, что

$$\|\xi\| \leq R \Rightarrow \|\Delta_2(\xi, e)\| \leq N$$

для всех $e$ и $\kappa$.

Объединяя модель объекта (5), адаптивную внутреннюю модель (15), (16), (17), (18), получим замкнутую систему

$$\begin{aligned} \dot{\chi} &= f(\chi) + g(\chi, e), \\ \dot{e} &= \kappa[A_o - B_{o2}\Delta_0(\xi, e)C_o]e + B_{o1}\Delta_1(\xi, e) + B_{o2}\Delta_2(\xi, e), \end{aligned}$$

где подсистема $f(\chi)$ имеет вид (10). Применяя аргументы, аналогичные [5], можно показать, что существуют предел насыщения $L$, сколь угодно малое $\epsilon > 0$ и число $\kappa^*$ такие, что при $\kappa > \kappa^*$ и $t \to \infty$ все траектории системы ограничены и

$$\lim_{t\to\infty} \| x_e(t) \| \leq \epsilon,$$

что соответствует выполнению цели (3).

**Моделирование.** В настоящем разделе рассмотрим моделирование задачи удержания заданного положения надводного судна в условиях действия внешних возмущений с неизвестными параметрами с использованием предложенного подхода.

Рассмотрим модель надводного судна (1) с параметрами

$$M = \begin{pmatrix} 3{,}0 & 0{,}0 & 0{,}0 \\ 0{,}0 & 2{,}0 & -0{,}5 \\ 0{,}0 & -0{,}5 & 1{,}0 \end{pmatrix}, \quad D = \begin{pmatrix} 1 & 0 & 0 \\ 0 & 1 & 1 \\ 0 & 1 & 1 \end{pmatrix},$$

возмущающие воздействия

$$\tau_d = \begin{pmatrix} \tau_{d1} \\ \tau_{d2} \\ \tau_{d3} \end{pmatrix} = \begin{pmatrix} 2 + 9\sin(0{,}75t + 9) \\ 1 + 3\sin(0{,}50t + 4) \\ 5 + 4\sin(0{,}25t + 7) \end{pmatrix},$$

вырабатываемые генератором (2) с параметрами

$$S_1 = \begin{pmatrix} 0 & 0 & 0 \\ 0 & 0 & 1 \\ 0 & -0{,}5625 & 0 \end{pmatrix}, \quad H_1 = \begin{pmatrix} 1 \\ 1 \\ 0 \end{pmatrix}^T, \quad w_1(0) = \begin{pmatrix} 2{,}0000 \\ 3{,}7091 \\ -6{,}1501 \end{pmatrix}$$

$$S_2 = \begin{pmatrix} 0 & 0 & 0 \\ 0 & 0 & 1 \\ 0 & -0{,}2500 & 0 \end{pmatrix}, \quad H_2 = \begin{pmatrix} 1 \\ 1 \\ 0 \end{pmatrix}^T, \quad w_2(0) = \begin{pmatrix} 1{,}0000 \\ -2{,}2704 \\ -0{,}9805 \end{pmatrix}$$

$$S_3 = \begin{pmatrix} 0 & 0 & 0 \\ 0 & 0 & 1 \\ 0 & -0{,}0625 & 0 \end{pmatrix}, \quad H_3 = \begin{pmatrix} 1 \\ 1 \\ 0 \end{pmatrix}^T, \quad w_3(0) = \begin{pmatrix} 5{,}0000 \\ 2{,}6279 \\ 0{,}7539 \end{pmatrix},$$

где $w(0) = (w_1^T(0) \quad w_2^T(0) \quad w_3^T(0))^T$.

Применим адаптивную внутреннюю модель (15), (16), (17), (18) с параметрами

$$F_i = \begin{pmatrix} 0 & 1 & 0 \\ 0 & 0 & 1 \\ -1 & -3 & -3 \end{pmatrix}, G_i = \begin{pmatrix} 0 \\ 0 \\ 1 \end{pmatrix}, \hat{\Gamma}_i^T(0) = \begin{pmatrix} 1 \\ 0 \\ 3 \end{pmatrix}, Q_i = \begin{pmatrix} 0 & 0 & 0 \\ 0 & 0{,}5 & 0 \\ 0 & 0 & 0 \end{pmatrix},$$

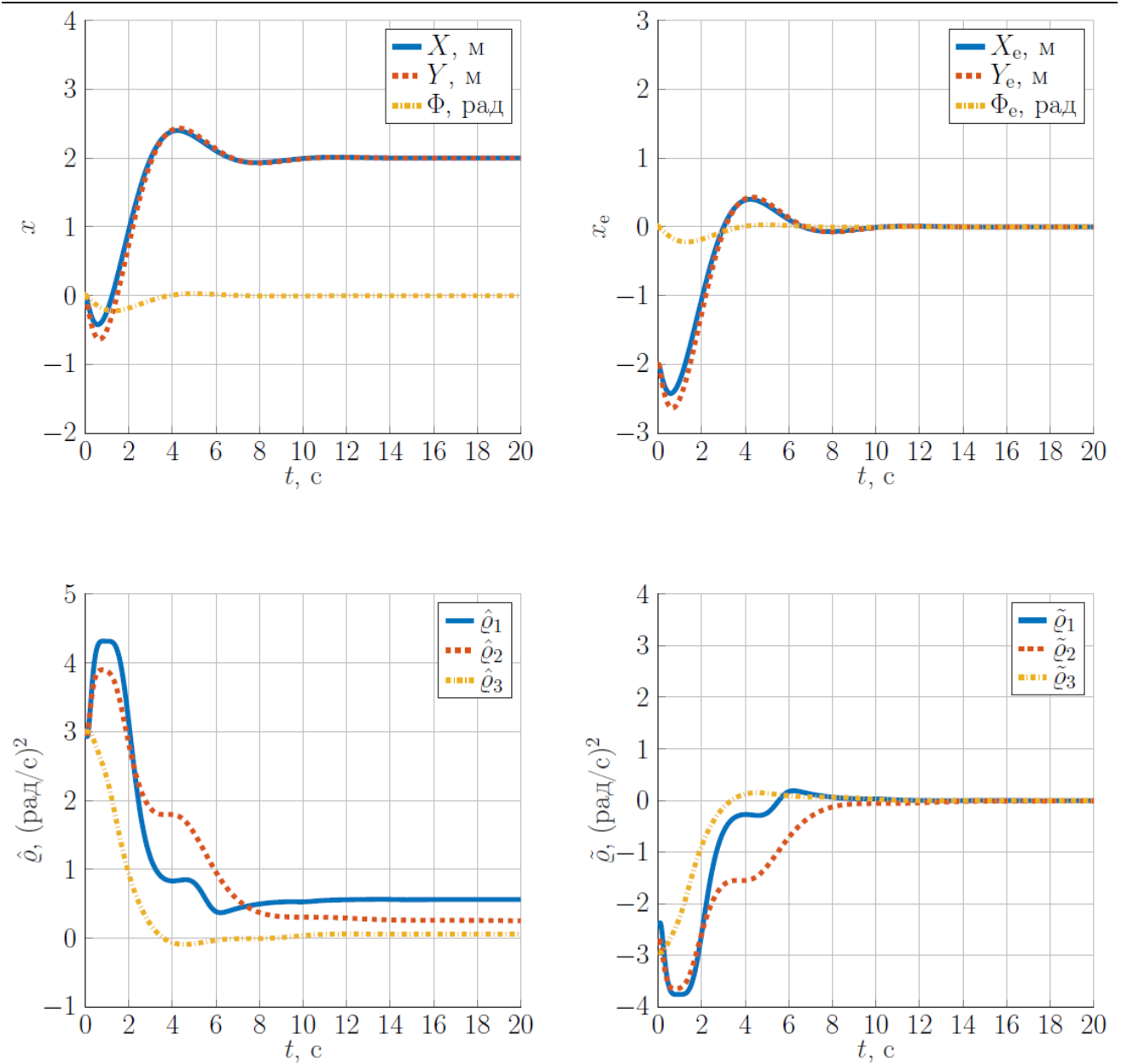

Рис.1. Результаты моделирования

$$L = 100, \quad \kappa = 100, \quad K_1 = K_2 = I_{3\times 3}, \quad M_0 = I_{3\times 3},$$

$$C_0 = C_1 = \begin{pmatrix} 30 & 3 & 3 \\ 3 & 30 & 3 \\ 3 & 3 & 30 \end{pmatrix}, \quad C_2 = \begin{pmatrix} 10 & 1 & 1 \\ 1 & 10 & 1 \\ 1 & 1 & 10 \end{pmatrix}.$$

Результаты моделирования задачи удержания заданного положения

$$x_r = \begin{pmatrix} X_r \\ Y_r \\ \Phi_r \end{pmatrix} = \begin{pmatrix} 2 \\ 2 \\ 0 \end{pmatrix}$$

приведены на Рисунке 1. На графиках показаны переходные процессы оценок $\hat{\varrho} = \{\hat{\varrho}_1, \hat{\varrho}_2, \hat{\varrho}_3\}$ неопределенных параметров $\varrho = \{\varrho_1, \varrho_2, \varrho_3\}$, содержащихся в блочных матрицах генератора (2)

$$S_i(\varrho_i) = \begin{pmatrix} 0 & 0 & 0 \\ 0 & 0 & 1 \\ 0 & -\varrho_i & 0 \end{pmatrix}, \quad i = \{1,2,3\},$$

а также — в силу свойства внутренней модели (6) о равенстве спектров матриц $F_i + G_i \Gamma_i(\varrho_i)$ и $S_i$ — в векторах

$$\Gamma_i(\varrho_i) = \begin{pmatrix} 1 & 3 - \varrho_i & 3 \end{pmatrix}, \quad i = \{1,2,3\},$$

а также соответствующие ошибки оценивания $\tilde{\varrho} = \{\tilde{\varrho}_1, \tilde{\varrho}_2, \tilde{\varrho}_3\}$. Как видно на графиках, ошибки оценивания $\tilde{\varrho}_i$ сходятся к нулю, а сами оценки $\hat{\varrho}_i$ стремятся к истинным значениям $\varrho_i$. В результате, как и ожидалось, ошибки регулирования $x_e = (X_e \quad Y_e \quad \Phi_e)^T$ сходятся к нулю, а выходные переменные $x = (X \quad Y \quad \Phi)^T$ стремятся к заданным значениям $x_r = (X_r \quad Y_r \quad \Phi_r)^T$.

**Заключение.** В работе предложение решение задачи удержания заданного положения модели надводного судна (1) в условиях действия параметрически неопределенных возмущений, описываемых с помощью (2). Предлагаемый подход предусматривает преобразование динамической модели (1) с учетом (2) к модели в форме (5), для которой синтезирован закон управления на основе адаптивной внутренней модели (15) и расширенного наблюдателя (16), (17), обеспечивающий достижение полуглобальной практической устойчивости системы и выполнение цели (3). Работоспособность предложенного подхода проиллюстрирована результатами компьютерного моделирования.

## СПИСОК ЛИТЕРАТУРЫ

ROBUST CONTROL OF A SURFACE VESSEL WITH ADAPTIVE REJECTION OF DISTURBANCES WITH UNKNOWN PARAMETERS

O.I. BORISOV, F.B. GROMOVA, A.Iu. ZHIVITCKII, A.A. PYRKIN

ITMO University, 197101, St. Petersburg, Russia

This paper solves the problem of station-keeping control of a surface vessel under conditions of sinusoidal disturbances with unknown parameters. The proposed control algorithm is based on the geometric approach with the use of the adaptive internal model and the extended observer. The paper analytically proves the boundedness of the trajectories of the system and their semiglobal convergence to an arbitrarily small set. The performance of the algorithm is illustrated by computer simulation.

Keywords: disturbance rejection, geometric approach, adaptive internal model, extended observer, surface vessel.